# Strain Effect in MgB$_2$/Stainless Steel Superconducting Tape

H. Kitaguchi*, H. Kumakura, and K. Togano

National Institute for Materials Science, 1-2-1, Sengen, Tsukuba 305-0047, JAPAN
* KITAGUCHI.Hitoshi@nims.go.jp

*Abstract*—The influence of mechanical strain on the critical current (I$_c$) is investigated for MgB$_2$/stainless steel (SUS316) superconducting tapes. The tapes are fabricated by using "powder in tube" method and deformation process without any heat treatment. The tensile axial strain along tape length is successfully induced to the sample by using a U-shape holder made of stainless steel (SUS304). Two samples are examined at 4.2 K in 5 T (B is applied perpendicular to the tape surface). While the initial I$_c$ at zero external strain state (I$_{c0}$) varies (30.4 and 33.3 A), normalized I$_c$ (I$_c$/I$_{c0}$) vs. external strain relations fall on the same curve. Linear increase of I$_c$ is observed from zero external strain state to 0.4% strain (107% of I$_{c0}$). Rapid and large degradation occurs at the strain exceeding 0.4 ~ 0.5%. High durability against stress can be expected for MgB$_2$/stainless steel superconducting tapes.

## I. INTRODUCTION

MAGNESIUM-DIBORIDE (MgB$_2$) superconductor [1] has been studied extensively by many research groups in the world since its discovery. Much effort has been performed to realize a long conductor that would lead to new possibilities of practical applications. Several groups reported conductor fabrication by using the "powder in tube (PIT)" method and heat treatment [2-5]. Recently, a new fabrication procedure by using PIT method without any heat treatment was developed by Grasso et al. [6]. This process is attractive because the process including no heat treatment can avoid the problems originating in high reactivity of Mg and B against sheath materials and/or atmospheric gas at high temperatures. It is also helpful to reduce production costs of conductors. We have reported MgB$_2$/stainless steel tape that carries high critical current (I$_c$) [7]. The transport critical current density (J$_c$) of the tape reaches 10,000 A/cm$^2$ at 4.2K and 5T. The effect of strain on I$_c$ is important in designing and fabricating any practical application. Thus, the investigation of the strain effect has been much expected in order to examine a feasibility of MgB$_2$ conductors. In this paper, we report I$_c$ of MgB$_2$/stainless steel tapes under tensile axial strain along tape length at liquid helium temperature, 4.2 K and 5 T.

## II. EXPERIMENTAL

The procedure of sample preparation has been reported so far [7]. Briefly, mono-core rod was fabricated with PIT method by using commercially available MgB$_2$ powder (Alfa-Aesar) and stainless steel (SUS316) tube. The rod was deformed into a tape by using groove-rolling and then flat-rolling machines. No heat treatment was performed in the tape preparation process. The final cross section of the tape was about 5.2 mm in width and 0.5 mm in thickness. The typical thickness of the MgB$_2$ core was about 0.2 mm. Samples of ~65 mm in length for the strain tests were cut from the tape.

The details of the strain test have been reported [8-11]. Sample setting and the principle of strain generation are summarized in Fig. 1. The U-shape holders were made of stainless steel (SUS304) with a surface coating of Ni. The sample was soldered by using typical Pb-Sn solder. The sample was fully fixed to the holder in order to deform unitedly with the holder. Current and voltage leads were soldered directly to the sample. Tensile strain can be induced by decreasing the distance between both ends of the holder. The holder was attached to the movement operated with a computer-controlled actuator and the strain can be controlled continuously in this apparatus. We calibrated the relation between the strain at the sample position and the move of the actuator for the holder by using strain gage at 77 K. In the calibration, the strain gage was fixed by using glue directly to the holder at the center of sample position instead of mounting a sample. The strain tests were performed at 4.2 K, 5 T (perpendicular to the tape surface). I$_c$ was determined from dc transport measurements with 1 µV/cm electric field criterion. The voltage was monitored at the center part of sample that corresponds to the gage position in the strain calibration. Current distribution to the holder itself was ignored in I$_c$ determination because the current contacts were put directly to the sample surface without touching the holder and the electric resistance of SUS304 in the magnetic field is enough high.

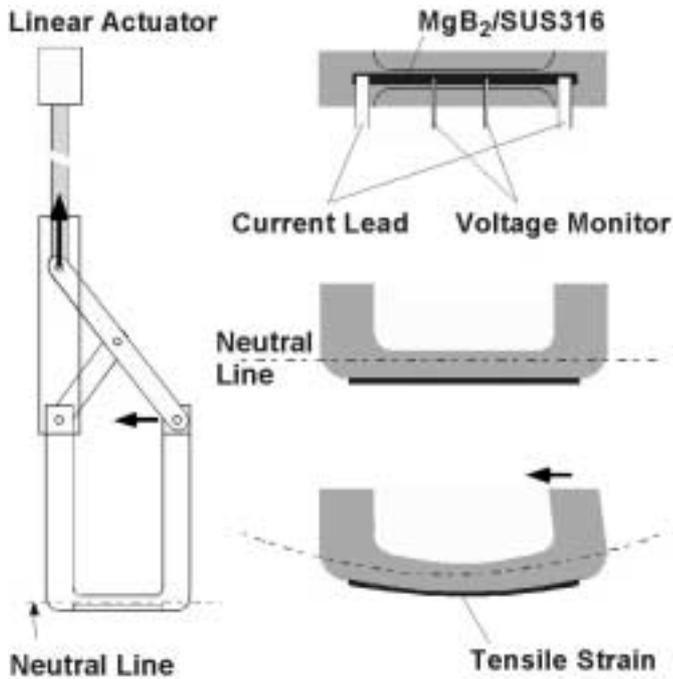

Fig. 1. Sample setting and the principle of strain generation. The sample and the holder deform unitedly and pseudo-uni-axial strain is induced in the sample.

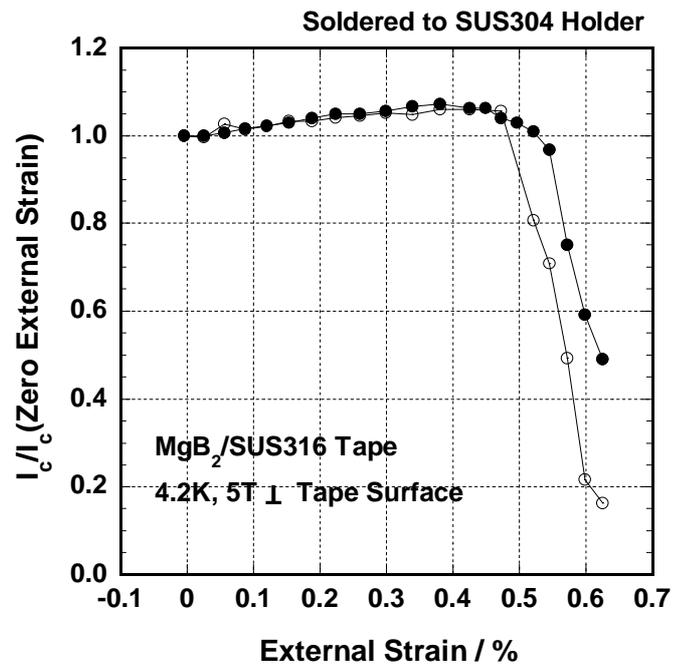

Fig. 2. $I_c$-external strain relation for $MgB_2$/stainless steel (SUS316) tapes at 4.2 K and 5 T. 0% external strain corresponds to "as-cooled" state of the setting in which the sample was soldered to SUS304 holder.

## III. RESULTS AND DISCUSSION

Results for two tensile tests at 4.2 K and 5 T are plotted together in Fig. 2. $I_c/I_{c0}$ ($I_{c0}$: $I_c$ value at zero external strain) was plotted against external strain in this figure. 0% external strain is defined to be as-cooled state of the test setting in which the sample was soldered to SUS304 holder. Tensile strain is defined to be positive strain. $I_{c0}$ values of the two samples were 30.4 and 33.3 A. The variation of $I_{c0}$ is supposed to originate in the inhomogeneity of the long tape. While $I_{c0}$ varies, $I_c/I_{c0}$ vs. external strain relations fall on the same curve with a good agreement. A linear increase of $I_c$ was observed from zero external strain state to about 0.4% strain (107% of $I_{c0}$). Rapid and large degradation of $I_c$ occurs at the strain exceeding 0.4 ~ 0.5%.

Pre-strain of the sample at zero external strain state can be caused by the difference of contraction between the sample and the holder due to cooling from soldering temperature to 4.2 K. This pre-strain is not clear because the thermal contraction of the tape has not been measured. The pre-strain, however, is thought to be small or negligible because the sheath material (SUS316) and the holder material (SUS304) are both stainless steel and have similar thermal contraction. The thermal contraction of SUS316 from room temperature to 4.2 K is estimated to be 0.31% [12]. The strength of the sheath material is supposed to be much higher than that of $MgB_2$ core, and then, a compressive inner-strain is supposed to be generated in $MgB_2$ core by the cooling (inner-strain means the strain at $MgB_2$ core). Therefore, the linear increase of $I_c$ will be explained by the decrease of inner-strain. Rapid and large degradation occurred at the strain exceeding 0.5% will be attribute to a fracture of the $MgB_2$ core. Assuming that pre-strain caused by the thermal contraction is negligible as discussed above, the $MgB_2$/SUS316 tape can be durable against tensile strain up to 0.4%. This value of critical strain (strain where rapid decrease of $I_c$ occurs) is twice higher than that for Bi-2223/Ag tape (0.2 ~ 0.3% at 77 K [11]). The critical strain of 0.4% corresponds to tensile stress of 0.8 GPa [13].

The results suggest that $MgB_2$/stainless steel conductors have a large possibility to be used in practical applications where the conductors are subject to large stress and/or strain. The toughness of the conductors is also helpful in the conductor production, winding, fabrication of application systems, and so on.

## IV. SUMMARY

The influence of tensile strain (along tape length) on $I_c$ was investigated for $MgB_2$/stainless steel (SUS316) tapes at 4.2 K and 5 T. The results are summarized as follows:

1. The strain tests were performed for $MgB_2$/ SUS316 tapes prepared by using PIT method without heat treatment. $I_c$-external strain relation was examined by using a U-shape holder made of SUS304 on which the sample was soldered.
2. A linear increase of $I_c$ was observed from zero external strain state to about 0.4% strain (107% of $I_{c0}$).
3. Rapid and large degradation of $I_c$ occurs at the strain exceeding 0.4 ~ 0.5%.
4. $I_c$-external strain curves for two samples fall on a master

curve with a good agreement.
5. The results show that MgB$_2$/stainless steel conductors have a large possibility to be used in practical applications where the conductors are subject to large stress and/or strain.